\documentclass[10pt, conference]{IEEEtran}

\usepackage{url}
\usepackage{graphicx}
\usepackage{epstopdf}

\usepackage{amssymb}
\usepackage{multirow}

\usepackage{times}
\usepackage{textcomp}
\usepackage{hyperref}

\def\ie{i.\,e.,~}
\def\eg{e.\,g.,~}

\title{ Simulating the Smart Grid  } 
\author{  
	\IEEEauthorblockN{Manfred P\"ochacker$^1$, Anita Sobe$^2$, Wilfried Elmenreich$^{1,3}$ }
	\IEEEauthorblockA{
	$^1$ Institute of Networked and Embedded Systems / Lakeside Labs \\
	Alpen-Adria Universit\"at Klagenfurt, Austria\\
	   $^2$ Institut d'informatique, Universit\'{e} de Neuch\^{a}tel, Switzerland \\
	   $^3$ Complex Systems Engineering, Universit\"at Passau, Germany \\
	   \\
	   manfred.poechacker@aau.at, anita.sobe@unine.ch, wilfried.elmenreich@aau.at
	 }

	}

\makeatletter
\def\blfootnote{
	\xdef\@thefnmark{}\@footnotetext
	}
\makeatother

\begin{document}

\setlength{\baselineskip}{.96\baselineskip}
\setlength{\abovecaptionskip}{0pt}

\maketitle

\blfootnote{ This work is supported by the Carinthian Economic Promotion Fund (KWF) under grant 20214/22935/34445 (Project Smart Microgrid Lab). 
We would like to thank Lizzie Dawes for proofreading the paper.}

\begin{abstract}
\setlength{\baselineskip}{.96\baselineskip}
Major challenges for the transition of power systems do not only
tackle power electronics but also communication technology, power market economy and user acceptance studies. Simulation is an important research method therein, as it helps to avoid costly failures. 
A common smart grid simulation platform is still missing.
We introduce a conceptual model of agents in multiple flow networks. Flow networks extend the depth of established power flow analysis through use of networks of information flow and financial transactions.
	We use this model as a basis for comparing different power system simulators.
	Furthermore, a quantitative comparison of simulators is done to facilitate the decision for a suitable tool in comprehensive smart grid simulation.

\end{abstract}

{ \bf Keywords} Smart Grid Simulation, Network Flow



\section{Introduction}
\label{sec:Introduction}

The power grid has started its transition towards the Smart Grid.
This development increases the level of complexity in the energy system \cite{Monti2010}, arising from the integration of distributed and renewable energy resources, smart meters, smart appliances and electric vehicles (EV), etc. into the electrical grid. 
For this reason, researchers of different fields are investigating a variety of topics related to the smart grid,
such as dynamic price markets \cite{Ramachandran2011}, demand response \cite{Albadi2008}, smart meters \cite{Depuru2011a} or prediction models \cite{Lu2010}.
Since the operation of a power grid is usually vital for its users, opportunities for testing novel approaches are very limited.
Therefore, to evaluate the impact of new methods for future smart grids, simulation has to be used.

	Most of the current simulators for smart grid scenarios originate in simulation of electric power systems, control circuits or agent based markets.
	Recently, development trends towards cross discipline simulations.
 	As the research field is broad \cite{Sobe2012}, it is hard to decide whether a particular simulator
 	fits the whole smart grid system or only parts of it.

	In this paper we answer the following questions regarding a simulator for the complete smart grid system:
\begin{enumerate} 
	\item What areas should a smart grid simulator cover? 
	\item Which functionalities do current simulators provide?
	\item How congruent are their results?
\end{enumerate}
To answer the first question, a generic model is introduced which is based on power system simulation and additionally considers two reliable circumstances. Firstly, the smart grid optimizes energy distribution through the use of more specific information and secondly, power markets are expected to manage the alignment of power demand with the current production through pricing.

The proposed model is based on a combination of an agent-based approach with electrical power flow, flow of data in the ICT network and flow of payments through the financial transaction network. 
After sketching the desirable state of smart grid simulation, we identify scopes of operation for several power system simulators available free of charge. 
Finally, four of the simulators are quantitatively compared by applying them to a common test case of power flow analysis.
This article provides an overview and results of a discussion on simulation design for smart grids and facilitates the decision of which tools and models to be used in future research.

The model including agents and flow networks is described in the next section. The third section contains a comparison of ten different simulators for power systems. Four of those are subjects of the quantitative comparison by means of a test-case in the succeeding part of the paper.
Finally, we give our conclusion.


\section{Agents in Multiple Flow Networks}
	\label{sec:NetworksOfFlow}

The overall goal of a smart grid is to contribute to greater efficiency, reliability and environmental sustainability in energy usage, which requires the best possible alignment of power generation, power consumption and (limited) storage capabilities.
Such an alignment impacts user behavior, cultural habits, social norms, power markets, climate conditions and many other factors.

\begin{figure*}
	\centering
		\includegraphics[width=\textwidth]{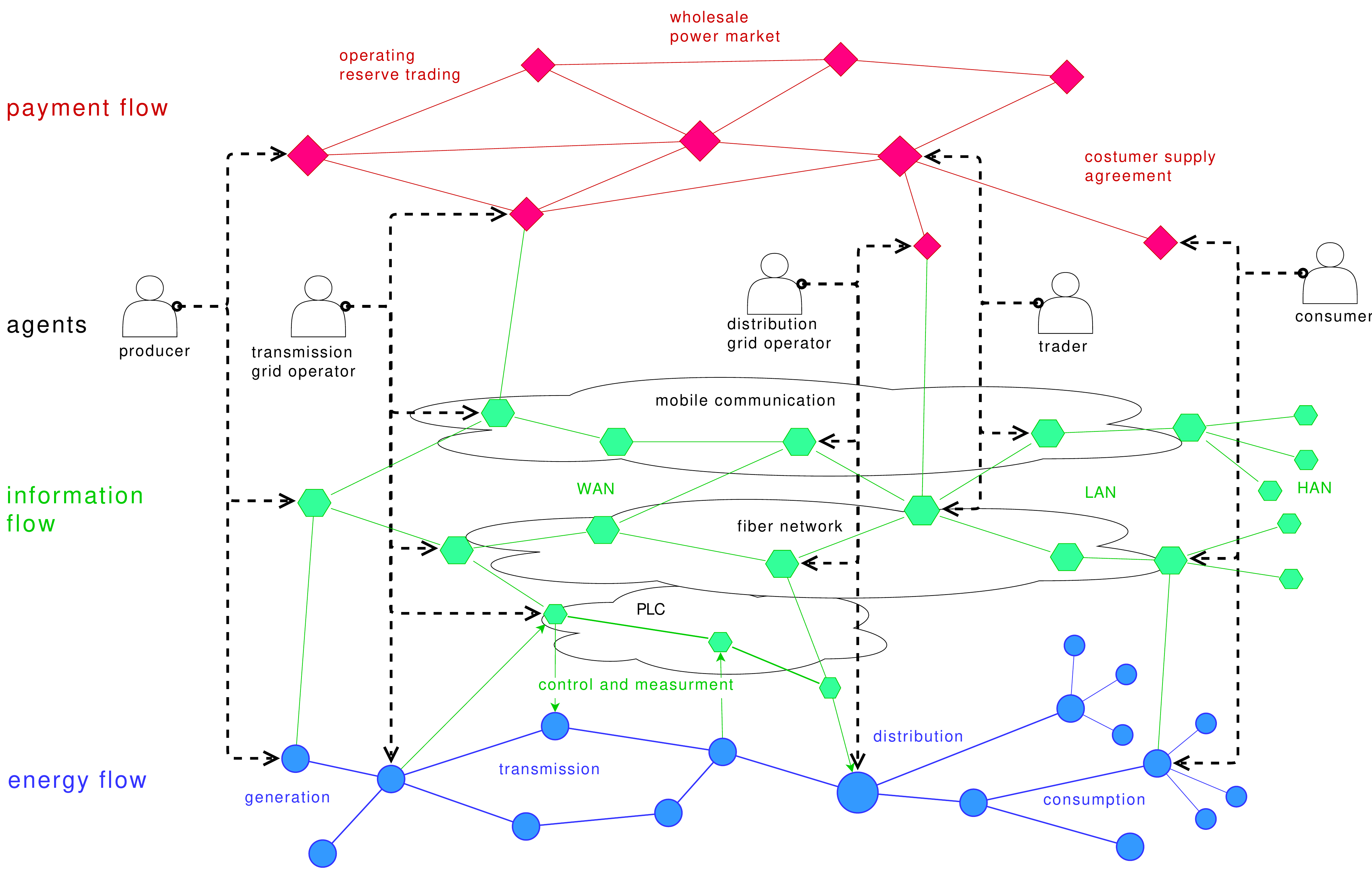}
	\caption{ A random smart grid example with agents in three flow networks (energy, information and payment). The nodes (circle for energy, hexagon for information, diamond for payment) in the network modify the flow; they are emitters and/or receivers of it. Nodes can have different functions with respect to their locations, \eg power generation, LAN routing, power market operator. The agents are the decision making entities and control several nodes and are the main interconnection between the different flow networks.  }
	\label{fig:FlowNW}
\end{figure*}

We propose a model that combines agent-based simulation and the notion of network flows.
A power system can be easily represented by a network model, as in power flow simulation. In the Smart Grid, this model is extended by: The flow of data in the ICT network and the flow of payment through the financial transaction network.
The agents are decision making entities that interact with each other through the three different types of networks: energy, information and payment flow networks.
The following example helps to explain this idea to the reader.

Let us imagine a simple smart grid scenario 
that includes the following entities: A consumer, a grid operator and a smart meter, an electric vehicle (EV) and a switching device.
Electrical power is transmitted from the smart meter (energized by the grid operator), through the switch into the storage of the EV. Information is transmitted by the consumer who controls the switch and connects the EV. The consumer is informed by the EV about its energy demand, by the smart meter about the current energy price and the switch about its state. Further information flows from the smart meter to the grid operator. 
The payment flow, (from the consumer to the grid operator) is based on billing information.
This simple example demonstrates that the three networks of flow, \ie energy, information and payment are tightly interwoven.
A node within the corresponding network modifies the flow and can be involved in a second or third type of flow network.
An agent is a super node that merges several nodes from different flow networks and tries to optimize its (the super node's) individual utility function.
The consumer in the example is an agent who controls the switch including the EV. The user also pays for the energy and receives information from various sources.  
The second agent is the grid operator. The operator is the recipient of the payment flow and controller of the smart meter.

Figure \ref{fig:FlowNW} depicts a more general example, which considers a random power system from generation to consumer. It includes five agents connected by the three flow networks.
The three networks are depicted by different node-symbols: circles for energy, hexagons for information and diamonds for payment flow.
Each agent is associated with its controlled nodes, marked by dashed lines. In this case not all nodes are associated with an agent. The agents form the main interconnection between the different flow networks. They use all their controlled nodes to maximize their utility function.
 Measurement processes are links from an energy node to the information network, functioning as an information source. 
 The opposite applies to control processes which use information.

This complex model of agents in multiple flow networks is generic.
It helps to structure and classify problems concerning the smart grid and can be scaled to any type of subsystem.
It is applicable on all layers; from intercontinental transmission lines to household devices.
The model can be extended by integrating other flow networks, \eg the gas distribution system.
It is also feasible to extend the model by including evaluation of measured data.

\subsection{The Agents and the utility function}

Each agent optimizes its utility function within its available scope of actions.
The utility function could be anything fitting properly into a mathematical formalism, for example, the reduction of costs and/or maximization of profit are common for agents in an economic environment. The optimization of the overall welfare is an alternative form of utility function.
The regular agents in a smart gird are consumers, producers, service providers, such as grid operators, and combinations of these roles. 
Within the proposed flow networks, the agents control a set of nodes and agents interact with each other through flow of any type.
Any type of social relation is not covered.

\subsection{Multiple Types of Flow Networks}

A complex network consists of a set of nodes interconnected by edges, also called \textit{links} \cite{Ahuja1993}. .
Any flow, defined as the amount of media transferred per unit of time, is a directed link connecting its emitter node with its receiver node. The weight of the link is given by the strength of flow, \eg the electrical power, data rate, etc.
According to the characteristics of the flowing media and the physical network apply different constraints which are used for calculations.

The flow, $f_X$ of type $X$, from a node $A$ to a node $B$, in a physical network is always limited by the capacity, $c$, of its infrastructure.
\begin{equation}
  f_{E,I}(A,B) \leq  c_{E,I}(A,B)
 \end{equation}
The index, $X$, includes all flow types for which the equation is valid, such as $E$, energy, $I$, information and $P$, payment.
If transmission is assumed to be lossless and the media is incompressible, the flow is antisymmetric.
\begin{equation}
 f_{E,P}(A,B)=-f_{E,P}(B,A)
\end{equation}
The total flow depends on the type of a node, i.e., whether a node is a source, a sink of flow or a transmission node. For a transmission node without any storage capabilities the total flow is zero.
\begin{equation}
 \sum_Bf_{E}(A,B)  \left\{ \begin{array}{ll}
	< 0 & \mbox{sink} \\
	= 0 & \mbox{transmission} \\
	> 0  & \mbox{source}
\end{array}
 \right.
\end{equation}
Further possible constraints are out of scope of this paper.

\subsubsection{Energy Flow in the Power Grid}
\label{sec:EnergyFlow}	
The principal measures in an AC grid are complex values of voltage and current. For scales above transient time, it is sufficient to deal with the active and reactive power (or the complex apparent power).
Network models are established in power systems engineering \eg in combination with Kirchhoff´s Law.
The \textit{power flow analysis} is based on them and is a common method to estimate the power distribution and voltage levels in a grid.
Modern power grid management uses much ICT infrastructure which itself relies on appropriate power supply.

\subsubsection{Information Flow over the ICT Infrastructure}
		\label{sec:InformationFlow}
		
		Information flow can be modeled as metadata containing a sender, a receiver and a transmission time. The only constraints refer to minimal time and maximal capacity for transmission. Any processor of information (or data) is a node in the network, identified by its address. Such generic models, also used in information theory \cite{Ahlswede2000}, typically consider logical connections between nodes, disregarding physical wiring.
		Nevertheless, it is much harder to capture information flow in a more coherent way than in terms of electrical power.
		One reason is that information is not well defined. Here, it is used very generally without definition of knowledge and data.  The information flow is defined by data transmission rate.
		The fact that data can be easily stored and replicated any number of times leads to different structures and flow dynamics of the information network. 
		The real ICT infrastructure is highly structured and based on many technologies diverging in data rates, used OSI layers, protocol standards, transmission quality and media. 
		
		 Many control circuits in the power grid keep it working properly. They are not part of an information flow network. A SCADA system forms a clear interface between power and information flow networks. In general, each measurement or control process connects the information flow with the power flow network and ICT usage is on progress in power system technology.

\subsubsection{Payment Flow at the Finance Transaction Network} 
		\label{sec:paymentFlow}

	 A node in the finance transaction network is an account that accumulates the payment flow over time.
	The payment flow results from transferring an amount of money from the sender's to the receiver's account. 	
	There is no physical transmitted media, so all the losses arise from finance transaction fees and the costs of accounting.
	The payment flow is antisymmetric and the nodes have no (theoretical) limit of storage capacity. The maximum transmission is limited by the state of the sender's account. The agent's utility function includes the state of the account.
		The legislative system gives constraints for bilateral contracts and energy market rules. Those agreements constrain all financial transactions. The payment flow is a simple and clearly defined way to track activities of and dependencies between agents, assuming all activities of power system agents happen in an economic context.
All financial transactions require secure ICT.

	
	\section{Smart Grid Simulation Software}
	\label{sec:SimulationTools}

\renewcommand{\arraystretch}{1.2}

\begin{table*}[t]
	\centering
		\begin{tabular}{c|ccccc|ccc|cc|cc|c|c}
				\hline			
  			& 	\multicolumn{5}{c|}{Power Flow} &\multicolumn{3}{c|}{Power Dynamics} &\multicolumn{2}{c|}{Information}& \multicolumn{2}{c|}{Payments} & \multirow{2}{*}{ABM}	& \multirow{2}{*}{Specials} \\
  			&  AC-PF & DC-OPF & OPF & CPF & 3P-PF & EMT & SA & TD & PLC & COM & MKT & DR    & 	 	&   \\
				\hline
				\hline
UWPFLOW		& \checkmark & \checkmark & 	&  	&   &  	& \checkmark	& 	& &  & & & & \\
TEFTS	  	&  & 	& 	&  & &\checkmark	&  &  \checkmark	& &  & & & &\\
\hline
MatPower	& \checkmark & \checkmark & \checkmark & \textasteriskcentered &   & & & & & & \textasteriskcentered & &  & \\
PSAT 			& \checkmark & \checkmark & \checkmark & \checkmark &   &  & \checkmark & \checkmark &  & & & & & GUI, GNE \\
IPSYS			& \checkmark & \checkmark & \checkmark &	 & & & & & & & & & & GNE \\
MatDyn 		&  & &  & & &\checkmark & \checkmark & \checkmark & & & & &  & \\
\hline
AMES   		& 	& \checkmark & & &  & & 	& 	& & & \checkmark	& \checkmark & \checkmark &  GUI \\
InterPSS  & \checkmark & \checkmark &	& & \checkmark & \checkmark & \checkmark & &  \checkmark & & \checkmark & & &  GUI, GNE \\
OpenDSS 	&  \checkmark & & & & \checkmark &\checkmark & \checkmark & &  \checkmark & \checkmark & & & &   GIC   \\			
GridLab-D	& \checkmark & & & & \checkmark & & \checkmark & \checkmark	&  \checkmark & \checkmark & \checkmark & \checkmark & \checkmark &Climate Data\\ 
				\hline
  \multicolumn{5}{r}  { \textasteriskcentered an optional package is available}
			
		\end{tabular}
	\caption{	Functionalities of ten freely available power system simulators. 	}
	\label{tab:FlowLayers}
\end{table*}

The power grid represents the current state of the system evolving towards the smart grid.
The simulation of electricity systems is a traditional field in engineering that distinguishes between dynamic and static power system analysis.
Today's simulation software typically performs various analyses within that field, as well as in related areas.
The software tools used originate either in simulation of power grids, control circuits or agent-based markets.

Commercial software developers therefore provide a variety of software packages. 
Those are mostly costly, highly specialized and hard to modify, thus are less suitable for research and/or teaching.
	We compare several freely available power system simulators, as listed by the IEEE Task Force on Open Source software for Power Systems\footnote{ http://ewh.ieee.org/cmte/psace/CAMS\_taskforce/software.htm }
	and assess their potential capabilities in terms of the agents in multiple flow networks approach. 

	\renewcommand{\arraystretch}{1.05}
\begin{table}
	\centering
		\begin{tabular}{|l|lr|}
		
   \hline
		\multicolumn{2}{|l}{Power Flow}  &\\
    \hline
		AC-PF & AC power flow &    \\
		DC-OPF & DC optimal power flow &    \\
		OPF & optimal power flow, the general AC version &\\ 
		CPF & continuous power flow &  \\
		3P-PF & three phase power flow for distribution grid & \\
    \hline
		\multicolumn{2}{|l}{Power Dynamics}   & \\
    \hline
		EMT & electro magnetic transients &\\
		SA & stability analyses, various types of &  \\
		TD & time domain simulation &  \\
    \hline
		\multicolumn{2}{|l}{Information Flow} &\\
    \hline
		PLC & programmable logic controllers  & \\
		COM & communication links   &  \\  %
		\hline
		\multicolumn{2}{|l}{Energy Market} &\\
    \hline
    MKT & energy market simulations &\\ 
    DR  & demand response, by consumers or producers  &\\ %
    \hline
		\multicolumn{2}{|l}{General and Special Functions} &\\
    \hline
    ABM & agent based modeling &\\ 
    GUI & graphical user interface & \\
    GNE & graphical network editor &  \\
    GIC & geomagnetically induced current & \\
		\hline

		\end{tabular}
	\caption{List of abbreviations of possible functionalities provided by power system simulators. They are listed and grouped as in Table \ref{tab:FlowLayers}.}
	\label{tab:Abrev}
\end{table}

Table \ref{tab:FlowLayers} contains a list of functionalities provided by ten different power system simulators.
We listed their major features and assigned them to different areas which are related to the three flow networks. The areas and the abbreviations of the functionalities are explained in Table \ref{tab:Abrev}.
The tables do not contain the simple DC-PF \cite{Purchala2005}.
The DC-OPF is commonly used for economical analysis to optimize welfare or benefit in accordance with a power system. As it is much easier to solve it is more frequently-used than the AC version of OPF \cite{Purchala2005}.
The CPF analysis is used to determine the maximum load condition of a grid. The 3P-PF calculates each phase of the power system instead of a aingle-phase equivalent circuit. Especially for asymmetric and two-phase loads the gained accuracy is worth the higher calculation effort.
All the power flow analyses are static or quasi-static, \eg in AMES where twenty-four (hourly) values within a day are used.
Faster processes, \eg transient simulation, are covered under Power Dynamics.
Within SA, several forms of stability analysis are considered, voltage and small signal stability as well as outage scenarios and short circuit simulation.
The PLC of GridLAB-D is simple programmable and assigned to the controlled device. The COM-links between PLC-objects are specified by reliability, bit rate and timeout.
All the energy market simulations provide different options for market rules which are not listed as separate functions in Table \ref{tab:FlowLayers}. 
Market simulations typically use ABM, which is a separate feature in Table \ref{tab:FlowLayers}.

The first two and oldest simulators listed in Table \ref{tab:FlowLayers}, UWPFLOW and TEFTS, are designed for static and dynamic power system analysis. 
The four simulators in the second group, from MatPower to MatDyn, are running in the Mathwork MatLab environment. 
MatPower\footnote{http://www.pserc.cornell.edu/MatPower} is a package of MatLab m-files intended as a simulation tool for researchers and educators. It is easy to use and modify from the MatLab-console as well as free and open source (apart from the required MatLab license).
MatDyn\footnote{http://www.esat.kuleuven.be/electa/teaching/matdyn} provides dynamic power system analysis and can be seen as a completion of MatPower.
IPSYS\footnote{http://www.ece.cmu.edu/\texttildelow nsf-education/software.html}  is a scripting tool used to define, manipulate, and analyze electrical power systems data of a prescribed format. A user can interact with single or multiple power system models through the IPSYS shell, a MatLab interface, or by using a GUI.
PSAT\footnote{http://www3.uclm.es/profesorado/federico.milano/psat.htm} is a GNU/Octave-based toolbox for electric power system analysis \cite{Vanfretti2007}. PSAT has further features, like Phasor Measurement Unit Placement, Eigenvalue Analysis, FACTS Models, Wind Turbine Models, and conversion of Data Files. It is one of the most complete freely available power system simulators \cite{Milano2005} with a comprehensive GUI and GNE in Simulink. 

The last four simulators in the list go beyond pure power system simulation. 
The AMES\footnote{ http://www2.econ.iastate.edu/tesfatsi/AMESMarketHome.htm }  Market Package is an extensible and modular agent-based framework for studying wholesale power markets. It is developed entirely in Java and provides output reports through table and chart displays. The hourly time steps and neglect of reactive power (\ie transmission losses) are weaknesses, but while the learning tool for agents is a clear innovation.
InterPSS\footnote{http://community.interpss.org} is an open-source project, mainly in Java, aimed at developing a simple to use, yet powerful software system for design, analysis, and simulation of power systems \cite{Zhou2007}. Its open and loosely coupled system architecture allows for easy import and export of components. 
Currently, the project seems ill-maintained.
The OpenDSS\footnote{http://sourceforge.net/projects/electricdss} is the open source Distribution System Simulator (DSS) of the Electric Power Research Institute in California. Its heritage is from general purpose power system harmonics analysis tools and is now a comprehensive electrical power system simulator \cite{Dugan2011}. Analyses are done exclusively in frequency domain. The GIC analysis is really special within the presented simulators. The program includes a scripting interface but it is mainly used through the provided dynamic-link library by other software, \eg MatLab.
GridLAB-D\footnote{ http://www.gridlabd.org} is a flexible simulation environment  that can be integrated with a variety of third-party data management and analysis tools \cite{Schneider2009}, \cite{Chassin2008}. The core algorithm coordinates the states of millions of independent devices. 
At its simplest, GridLAB-D examines in detail the interplay of every part of a distribution system with every other. GridLAB-D does not require the use of reduced-order models for the aggregate behavior of consumer or electrical systems. The import module for climate data is a unique feature. The scripting language is particularly designed for that software.

The development of the ongoing smart grid transition also affect simulation software as noticeable in Table \ref{tab:FlowLayers}.
The distribution system is greater detailed by recent simulators \cite{Schneider2009}.
The modeling paradigm changes and well-defined differential equation models get enriched, \eg by agent based modeling, stochastic models or game theoretic approaches \cite{Bompard2005}.
The simulators interconnect with other disciplines, primarily to economics but also to ICT systems \cite{Nutaro2007} - implementation of other forms of energy or climate data 
are other further examples.
The code of recent simulation software is more modular and easier to integrate into other programs. 
In some cases, researchers still design their own software, especially for development of alternative methods or for very new and specific problems.
The Mosaik \cite{Schuette2011,Schuette2012,Schuette2012a} simulation framework\footnote{http://srvevm01.offis.uni-oldenburg.de} is an example where only existing simulators are recombined to a new simulation framework. 
 An advanced GUI makes the simulators accessible for researchers of different fields.

\section{Application on a Test Case}

\begin{figure}
	\centering
		\includegraphics[trim = 3mm 0mm 0mm 0mm, clip, width=0.5\textwidth]{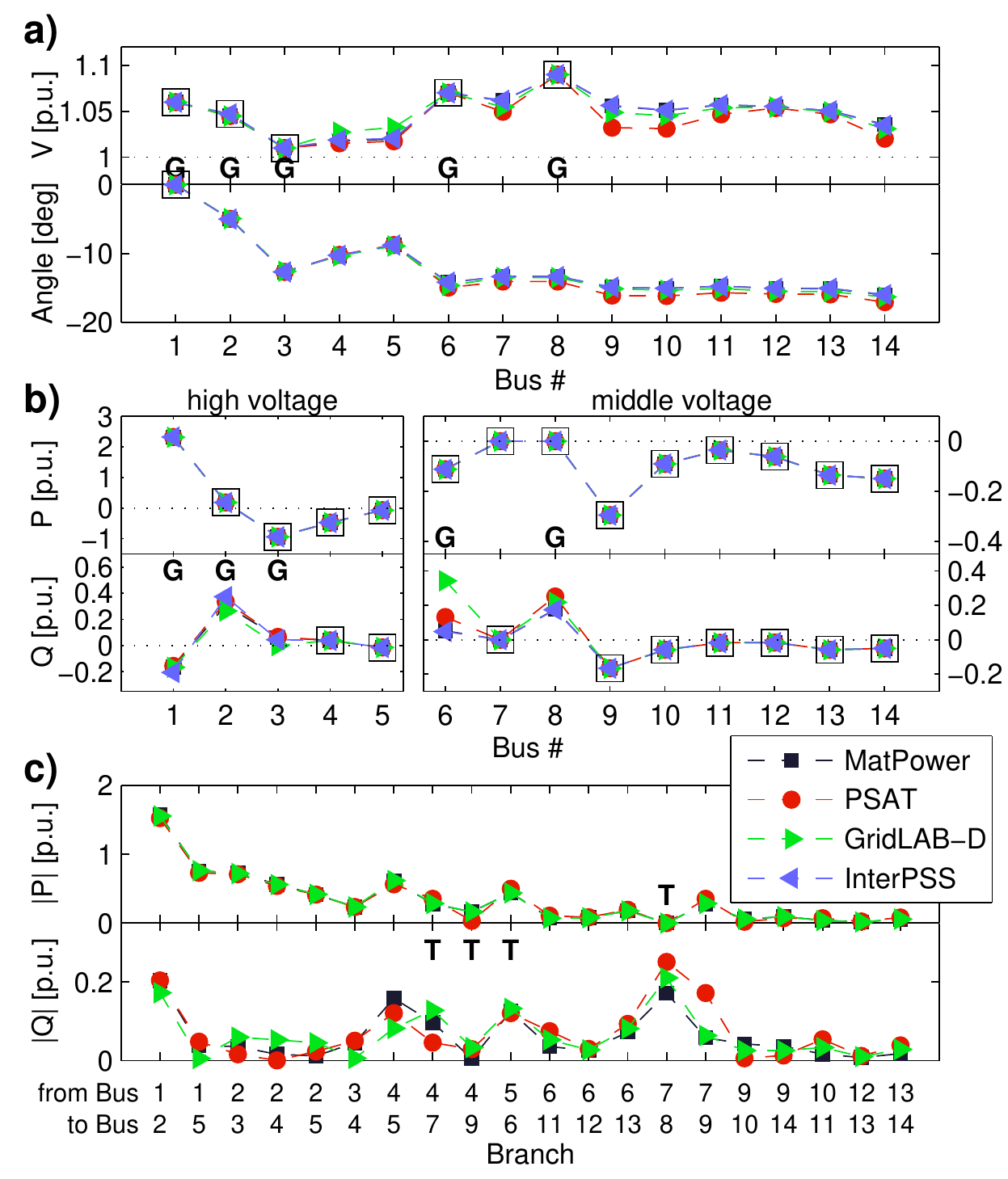}
	\caption{  
Solutions for the power flow of to the IEEE 14-bus test case from four different simulators. The black squares mark the input data for the analysis. Figure \ref{fig:PowerFlowResult}a shows the values for bus voltages as multiples of the nominal voltage and the voltage angles.
	Figure \ref{fig:PowerFlowResult}b illustrates the real and reactive power at each bus in units of $100 MW$ or $Mvar$ respectively.  
	In Figure \ref{fig:PowerFlowResult}c, the active and reactive power flow through each branch is shown.	
	$T$ marks a transformer in the power line, $G$, a generator in the bus. The legend applies to the whole figure. }

	\label{fig:PowerFlowResult}
\end{figure}

After the overview on the range of functionalities of the different power system simulators, we apply four of them to the same problem and compare the results. 
We selected four of the simulators from Table \ref{tab:FlowLayers}: MatPower, PSAT, InterPSS and GridLAB-D.
We used the IEEE 14-bus test case to perform power flow analysis.  This exemplary power system scenario is publicly available\footnote{http://www.ee.washington.edu/research/pstca/} in the IEEE common data format.
The input files exist for many simulators, including MatPower\footnote{ in the package-download \textit{case14.m} }, PSAT\footnote{ the path \textit{tests/d\_014\_mdl.m} in the package} with Simulink and GridLAB-D\footnote{ in \textit{gridlabd-course-1\_1/Solutions/4.1/4\_5\_IEEE\_14-bus\_Network.glm}}. The provided files have been used for the simulation. 
The demonstrated InterPSS results originate from the InterPSS \textit{Loadflow Study Guide}\footnote{http://community.interpss.org/Home/user-gude/loadflow-user-guide}.
 This test case is frequently used in studies \cite{Krumpholz1980,Milano2005,Monticelli1985}.
Five of fourteen buses are at high (nominal value $69kV$) and nine at middle voltages (nominal value $13.8kV$, one at $18kV$).

Figure \ref{fig:PowerFlowResult} shows the results of the power flow analyses of the four simulators. The top two charts show the four bus variables, the predefined ones in a black square. Busses with generators are marked with a bold 'G', branches containing a transformer with a 'T'.
In Figure \ref{fig:PowerFlowResult}a the bus voltages (in multiples of the nominal value) and voltage angles are shown.
The PSAT angle values of the middle voltage buses are slightly below the matching others.
Figure \ref{fig:PowerFlowResult}b depicts the total active power, $P$, and reactive power, $Q$, for all busses. The high and middle voltage busses are separated and have different scales. The convention for shown data is power producers with positive and consumers with negative sign. 
Deviating values can only be seen for the reactive power of bus six. The power values for load busses are given.
Figure \ref{fig:PowerFlowResult}c shows the absolute values of transmitted active and reactive power in each branch. For InterPSS, no data are available (see conclusion). GridLAB-D provides only the complex line current. The lines transmitted power, $S_{ij}$ is calculated from the From-Bus-Voltage and the conjugate line current according to: $ S_{ij} = -U_i \cdot I^*_{ij} $.
 The branches are in order of ascending bus numbers. 
While active power results are quite consistent, those for reactive power show the highest variation between the simulators.

The results of the four simulators match quite well in average, but nevertheless, there are marked differences in some areas.
The provided files and used models are not absolutely equal which could cause the differences at specific points. We found the following examples: Phase shift of transformers is not included in all simulators and PSAT assumes five degrees at the branch from bus four to nine. An additional reactive load is included at bus nine of the GridLAB-D model. The PF calculation needs an iterative solver, most programs use Newton-Raphson while GridLAB-D uses the Gauss-Seidel method.
The branches transmitted power is calculated from the analysis results and calculation methods for this differ.

%
%
\section{Conclusion}
\label{sec:Conclusion}

With the model of agents in multiple flow networks, we spanned a framework for a more complete view on the complex topic 'smart grid' \cite{Monti2010} and its simulation.
 This principal model is suitable for extension and more detailed itemization, adaptable (technically both) to specific cases and helpful in simulation and model building.
 We used it as a benchmark to compare ten free power system simulators, whereof four are also quantitatively compared.

Six simulators deal mainly with power system problems in different time domains. Functions for ABM or information and payments flow are not supported in those simulators. PSAT is the most complete of them, including a comprehensive GUI and chart display functions and would be a suggested first choice for power system engineering students. Implementation in a wider simulation environment seems possible with the command-line version. MatPower could work as a simpler  power system component in a broader, MatLab based simulation platform.

The other four, which are the newer ones, come with information- and/or payment-flow features.
AMES is only suitable for market simulation. User model adaption and expansions to information and energy-flow would need extensive Java development.
InterPSS might be comprehensive through the provided functions, open structure, the GUI and support of user models. Unfortunately, the source does not seem to be up to date and we could not run it. 
OpenPSS provides good functionalities in all fields except market simulation and is easy usable by other software. With the program download comes extensive documentation including examples.  
The development of GridLAB-D is progressing. Its current state is illustrated by the tutorial slides (including exercises). The distributed modeling approach makes it different but the software can match up to traditional power system analysis. All model coding and data management is, until now, in the user's responsibility; users are expected to be researchers or developers.

Finally, the choice of the simulator depends on the research questions to be answered and possible existing simulation tools to combine with.
The comparison of operation area and power flow analysis simplifies the selection of a proper research tool for smart grid simulation.
Engineers as well as economists, politicians and social scientists strive for their, maybe different, approaches.


\bibliography{SmartGrid}

\bibliographystyle{ieeetr}
%




%

\end{document}